# Preprints as accelerator of scholarly communication: an empirical analysis in Mathematics

Zhiqi Wang[1,2], Yue Chen[1], Wolfgang Glänzel[2,3]

**Abstract**

In this study we analyse the key driving factors of preprints in enhancing scholarly communication. To this end we use four groups of metrics, one referring to scholarly communication and based on bibliometric indicators (Web of Science and Scopus *citations*), while the others reflect usage (*usage counts* in Web of Science), capture (*Mendeley readers*) and social media attention (*Tweets*). Hereby we measure two effects associated with preprint publishing: publication delay and impact. We define and use several indicators to assess the impact of journal articles with previous preprint versions in arXiv. In particular, the indicators measure several times characterizing the process of arXiv preprints publishing and the reviewing process of the journal versions, and the ageing patterns of citations to preprints. In addition, we compare the observed patterns between preprints and non-OA articles without any previous preprint versions in arXiv. We could observe that the "early-view" and "open-access" effects of preprints contribute to a measurable citation and readership advantage of preprints. Articles with preprint versions are more likely to be mentioned in social media and have shorter *Altmetric attention delay*. Usage and capture prove to have only moderate but stronger correlation with citations than Tweets. The different slopes of the regression lines between the different indicators reflect different order of magnitude of usage, capture and citation data.

## 1. Introduction

Preprints play an increasingly important role in scholarly communication in many fields, notably Physics, Mathematics and Computer Science. More recently, researchers in the life sciences turn to embrace preprint publishing: The corresponding subject preprint repository, bioRxiv, was launched in 2013. In addition, the open science initiative with open access publishing as one of its main aspects has already promoted the preprint development and made preprints attract more attention than ever (Berg et al., 2016; Chiarelli, Johnson, Pinfield, & Richens，2019). While there are still debates on the value of such kind of un-refereed scholarly manuscripts (i.e. preprints) (Teixeira da Silva, 2018; Rawlinson & Bloom, 2019), the growth of preprint publications as well as preprint services is fast in recent years (Wang, Glänzel, & Chen, 2018; Narock & Goldstein, 2019). Furthermore, in the report of 2019 Altmetric Top 100 released by Altmetric.com on 17 December, 2019 (cf. Altmetric, 2019), which claims to highlight the 100 most-discussed works of 2019, four arXiv preprints are included in the list, and noticeably one arXiv preprint is ranked the first as the most discussed paper of 2019. All of these indicate that the landscape of preprint publishing is evolving

Corresponding author: Zhiqi Wang
E-mail: zhiqi_wang90@126.com
ORCID: 0000-0003-2216-4244

[1] WISE Lab, Dalian University of Technology, Dalian 116024, China
[2] ECOOM and Dept. MSI, KU Leuven, Naamsestraat 61, Leuven, 3000, Belgium
[3] Library of the Hungarian Academy of Sciences, Dept. Science Policy & Scientometrics, Budapest, Hungary

rapidly and more researches are necessary in order to reveal and validate the driving factors of preprints as accelerator of scholarly communication and their impact within academia and broader society.

Preprint publishing is primarily intended to convey most recent research results to the relevant target group in a highly efficient and fast way and considered as an important step towards a more open and transparent peer review process (Ginsparg, 2011).The main benefits of preprints are recognized as early discovery, fast and wide dissemination, free access and early feedback (Chiarelli et al., 2019). While the studies focusing on measuring these advantages are not yet sufficient and the full impact of preprint publishing remains to be seen. In this context also the question arises of what the added-value of supplementing traditional citation-impact indicators by new metrics, which are more related to research impact reaching out beyond scholarly communication (Glänzel & Gorraiz, 2015), may be.

A typical function of preprints is considered to bridge the time gap between the preparation of a manuscript and its publication in a journal (Ginsparg, 2011; Wang, 2019). Björk and Solomon (2013) reported that the publication delay (i.e. the time lag between the date of a paper being received by a journal and its print publication date) of mathematics journals was 13.3 months on average, longer than several other disciplines in the sciences, for instance Chemistry (8.9 months) and Physics (10.9 months), which is an important factor affecting Mathematicians heavily relying on preprints for sharing and tracking new research findings and ideas (Fowler, 2011). The primary purpose of preprints is to speed up and enhance communication with the community and there are practically no limitations with respect to the covered time frame, that is, revised, accepted or even formally published journal articles may also be uploaded to the preprint server, e.g., arXiv, during or even after the peer-review process. However, the time elapsed between the arXiv submission/update dates and the different time stages in the reviewing process of their journal version reflecting the interactive stages of the manuscript's evolutionary process is often not sufficiently clear. However, their knowledge is important for understanding the role of preprints in speeding up the dissemination of research results.

Furthermore, another important benefit of preprint publishing regards the enhancement of visibility, the valuable feedback by the community prior to final publication and the potential increase of impact. Bibliometrics with its proven tools for the measurement and evaluation of the impact of traditional journal articles could also be used for the assessment of the impact of preprints. Citations in scientific literature reflect the use of information within the framework of scholarly communication and are therefore used to "metrically" support research assessment (Glänzel & Schoepflin, 1999). Although citations can not describe the totality of the reception process, they give, according to Glänzel and Schoepflin (1999), "a formalised account of the information use and can be taken as a strong indicator of reception at this level".

While it is widely believed that depositing manuscripts in an open-access repository, institutional or subject-based archive could increase the number of citations received later by an article (Brody et al., 2004; Kurtz et al., 2005; Davis & Fromerth, 2007; Moed, 2007; Larivière et al., 2014; Mikki, 2017; Wang et al., 2018; Piwowar et al., 2018), bibliometric research has found controversial answers on the question of whether there is a citation advantage of publishing open access over toll access (Craig, Plume, McVeigh, Pringle, & Amin, 2007; Kurtz & Henneken, 2007; Davis, 2011; Moed, 2012). The biggest challenge in the citation analysis of preprint publishing is that it is difficult to establish any causality in the citation advantage of preprints. Kurtz et al. (2005) formulated three factors, "Open Access", "Early Access" and "Self-selection Bias" as the possible explanations for the effect of preprints on citations, and suggested that in most cases, they were non-exclusive and had combined effect. However, additional factors might also exist, as indicated by Henneken et al.

(2006), such as the increasing maturity and perfection of electronic publishing and academic search engines. For example, Google Scholar and Microsoft Academic enable documents on preprint servers to be immediately retrieved at the time of deposition and thus grant free access to readers worldwide, promoting large global visibility.

Nonetheless, seemingly obvious causal relations are not at all straightforward as has been shown by Glänzel and Heeffer (2014) in the context of the possible effect of downloads on citations, where the measured effects were rather bidirectional: Frequently cited papers were often downloaded before citing but highly cited papers, in turn, were frequently downloaded as well because of their popularity and interest found in the community, and because they had become a must-read in the respective research field. This is the reason why we are not aiming at finding any causal relationship in our research. We rather try to shed light on how preprint publishing would possibly reshape the publication behaviour of researchers and affect the readership and impact of their research outputs, if the scientific community, like Mathematicians, has widely acceptance and incorporate of preprint culture.

The emergence of online scholarly communication with different forms of fast and free access for broad user communities resulted in the demand for new and alternative indicators for capturing the variety of information use and impact within and beyond the framework of traditional scholarly communication (Glänzel & Chi, 2016). *Usage metrics* and *altmetrics* are the most known supplements to citation impact (Glänzel & Gorraiz, 2015). Brody, Harnad, & Carr, (2006) studied the correlation between downloads and citations of arXiv e-prints and found that the short-term Web usage impact of arXiv preprints predicted a medium-term citation impact of their final versions published in journals. Shuai, Pepe, and Bollen, (2012) found that there was a statistically correlation between the earlier Twitter mentions of arXiv papers and their downloads from the arXiv server and citations in Google Scholar. The *usage count* reported by Web of Science (WoS) shows the interest of the users of the database for further information, providing a different perspective of knowledge transfer (Wang et al, 2016; Chi & Glänzel, 2017; 2018). Positive correlations with *usage counts* and *citations* in WoS in several disciplines for journal papers have been detected (Chi & Glänzel, 2017; 2018), however little has been done so far to study the usage pattern of preprints in the context of the relationship between usage counts and citations.

*Altmetrics* are metrics for measuring diverse groups of actives on social web platforms, aiming to give new insights to the assessment of wider societal impact of research outputs that is invisible through traditional citation-based indicators (Priem & Hemminger, 2010; Bornmann, 2014). Compared with other altmetric indicators, Mendeley readership seems to be the most promising indicator as a supplement to citation impact of preprints due to several factors:

(1) Mendeley readership has a significant positive correlation with citations in most disciplines (Haustein, Larivière, Thelwall, Amyot, & Peters, 2014; Thelwall & Sud, 2016; Zahedi, Costas, & Wouters, 2017), and the readership profiles provided by Mendeley in terms of country, research area and academic status reflect a broad scale of knowledge dissemination (Mohammadi & Thelwall, 2014)
(2) The coverage of Mendeley is distinctly better than that of other web sources (Haustein et al., 2014) and retrieval from Mendeley by arXiv identifiers is supported.
(3) Last but not the least, Mendeley readership is also a faster and better indicator of early impact (Thelwall, 2017; Maflahi & Thelwall, 2018; Thelwall, 2018), which is quite important for assessing preprints impact.

Several studies analysed the impact assessment of preprints based on online readership. Bar-Ilan (2014) found that 47% of the publications of 100 European astrophysicists indexed in Scopus were in arXiv, whereas 40% arXiv papers had been covered in Mendeley, higher than the proportion of Scopus publications in Mendeley (27%). A recent impact study of bioRxiv preprints showed that 96.3% of journal papers deposited in bioRxiv were covered in Altmetric.com and received 75% more Mendeley readers than non-deposited papers (Fraser, Momeni, Mayr, & Peters, 2020). Similar result has also been found in preprints published in three main journals in Library & Information Science (LIS) (Wang, Glänzel, & Chen, 2020): 97% arXiv papers were covered in Mendeley and enjoyed a significant readership advantage compared with non-arXiv papers. Yet according to Zahedi et al. (2017) the presence and density of Mathematics journal articles in Mendeley is much lower than that in other disciplines. Therefore it is necessary to explore whether and in how far the existence of preprint versions affects the readership impact of their journal counterparts in Mathematics and how the relationship with the citation impact takes shape.

Twitter is another widely discussed social web source for altmetric analysis (Thelwall, Haustein, Larivière, & Sugimoto, 2013; Costas, Zahedi, & Wouters, 2015). Compared with Mendeley, Twitter is more widely used outside academia (Thelwall & Kousha, 2015). Although the coverage and density of scientific publications in Twitter are very low in many disciplines, especially in Mathematics (Haustein, Costas, & Larivièr., 2015), which challenges the reliability of the indicators based on Tweets (Costas et al., 2015), they provide new and interesting insights into the diverse and wider society impact of research (Sugimoto, Work, Lariviere, & Haustein, 2017), which is particularly important for understanding the added-value of preprints to traditional journal literatures. Besides, although the role of social media in supporting the discoverability and perception of preprints has been emphasized (Chiarelli et al., 2019), at present there is little evidence on how and to what extend the social media would work. In order to fill this gap, we conduct an altmetric analysis of the social attention impact of preprints in Mathematics through two indicators: the number of *Tweets* and *Altmetric 'first seen date'* of the two versions of preprints (i.e. arXiv versions and journal versions), and compared with those of non-OA papers without any previous preprint versions in arXiv.

Mathematics, as one of the fields where a long-standing culture of sharing preprints with members of the scientific community exists, with arXiv being by far the most widely used platform, offers a unique environment to study the current issues around preprint publishing due to several factors. First, Mathematics has a higher share of WoS papers deposited in arXiv, compared with other fields in science, which is even higher than that in Physics (Larivière et al., 2014). What's quite remarkable that the share of the number of arXiv submissions in Mathematics of the total arXiv submissions in 2018 (22.4%) has already been higher than those in Condensed Matter (15.2%), High Energy Physics (15.0%) and Astrophysics (14.8%) (cf. ArXiv, 2019), the top three sub-disciplines with the largest amount of arXiv submissions in Physics. Second, the ongoing process in favour of green OA in Mathematics has increasingly attracted attention, and self-archiving via arXiv is regarded as the key component to realise OA publishing in Mathematics (Ginsparg, 2016; Müller & Teschke, 2016; Bannister & Teschke, 2017). About 40% of publications indexed the Zentralblatt MATH (zbMATH, https://zbmath.org/) – the world's most comprehensive and longest-running abstracting and reviewing service in pure and applied mathematics – has been linked to their arXiv versions for the main subjects according to the Mathematics Subject Classification of the American Mathematical Society and the share is relatively stable in the last few years (Teschke, 2018). By providing insight into impact of preprints, this research contributes to the clarification of the role of preprint publications in scholarly and scientific communication in a broader sense, and sheds light on three main questions:

RQ1. How large is the publication delay of preprints in Mathematics and how it would affect the preprint publishing behaviour of authors?

RQ2. What are the characteristics of the aging patterns of arXiv versions of preprints and what's the relationship between the citations to preprint versions with the citations to their journal versions?

RQ3. Do previous preprint versions have measurable effect on the impact of their journal versions and is this different from those without previous preprint versions?

In order to answer these questions, we follow several key steps. *Source items* have been retrieved from the Clarivate Analytics WoS Core Collection (for the journal articles assigned to three core sub-disciplines in Mathematics) and the arXiv e-print archive. Articles and preprints are matched using a semi-automated procedure. Also citations received by both preprints and articles have been retrieved from the WoS database. Scopus citations and altmetric indicators could be added via DOI or arXiv ID matching. The detailed description of the procedure is given in the following section.

## 2. Data sources and processing

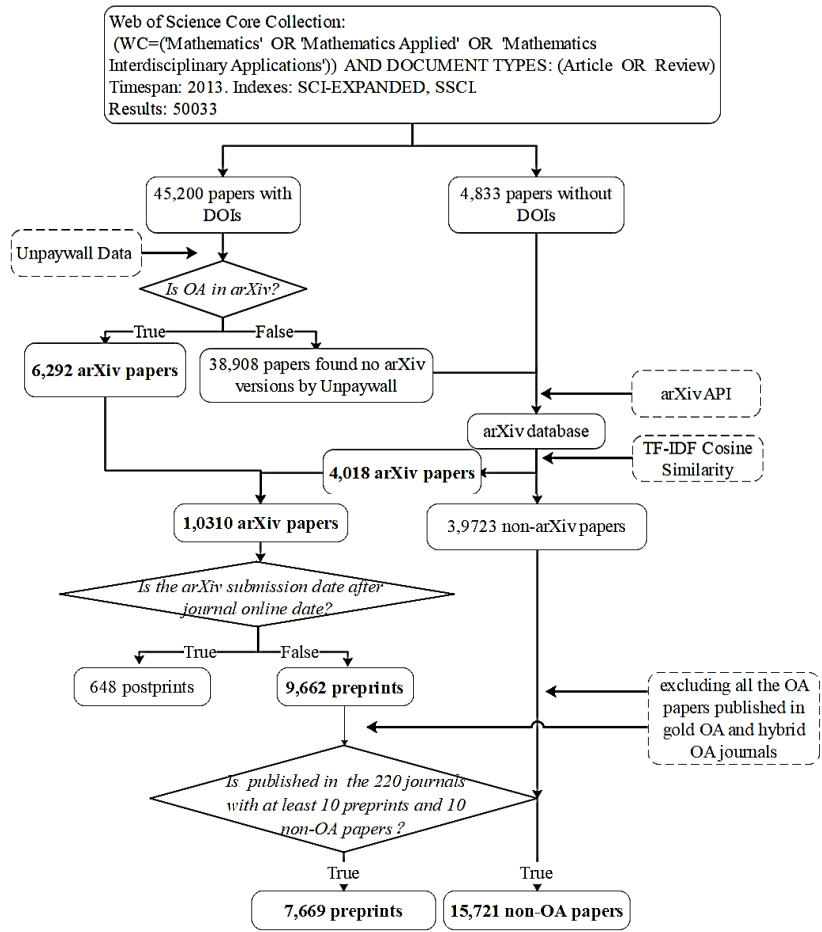

**Fig. 1.** Process for matching WoS papers with arXiv papers and sample data selection

For this study, two data sources are used, the arXiv e-print archive and WoS Core Collection database, and the process for matching WoS papers with arXiv papers and sample data selection is shown in Fig. 1. Each of these process steps is described in detail as follows:

*In the first step*, all papers (document type: "article" and "review") assigned to the three WoS Subject Categories 'Mathematics', 'Mathematics Applied', and 'Mathematics Interdisciplinary Applications' (denoted by "M1", "M2" and "M3" in the following), and indexed in the Science Citation Index Expanded (SCIE) and Social Sciences Citation Index (SSCI) of WoS database with the publication year 2013 are selected for the study. The number of documents amounts to 50,033, including 49,853 articles and 180 reviews. We have chosen the publication year 2013 for two reasons: a. Information in Mathematics is aging slowly as measured by citations (e.g., Rousseau, 1988; Zhang & Glänzel, 2017) and 5-year time window could provide a sufficiently large citation window (cf. Glänzel, Thijs, & Chi, 2016), and b. the publication year 2013 allows collecting information on usage and altmetrics alongside the citations.

*In the second step,* after having downloaded the necessary bibliographic data, metadata and citations, Unpaywall API is used to determine whether access is available in arXiv. In this step, 6,292 papers are found with OA versions deposited in arXiv. Since Unpaywall uses oaDOI data to find OA and the recall of the system is estimated about 77.0% (Piwowar et al., 2018).

*In the third step,* similarity matching is used to match the remaining 43,741 papers via arXiv API, including 4,833 papers without DOIs and those 38,908 papers with DOIs that could not be matched for any reason with its arXiv version in the second step. In order to increase recall we proceeded in the following two steps. First, two text-based links are established: a. direct correspondence between the arXiv titles and WoS titles; b. fuzzy matching between arXiv and WoS titles and first authors. Second, each link-pair found in the first step is validated through the tf-idf cosine similarity computed between their titles, abstracts, and authors respectively. The matching process is implemented in Python and the similarity score of each matching pair is computed. In order to determine an appropriate threshold of the scores ($S_{min}$) that the matching results have to exceed to be accepted, we take the 6,292 matched items found by Unpaywall as the evaluation dataset, which is quite reliable since the precision of the Unpaywall system is estimated about 96.7% as reported by Piwowar et al. (2018) and 99.1% by Archambault et al. (2014). In this case, it comes out that $S_{min}$ for the titles, abstracts and authors is about 0.8, 0.7 and 0.6 respectively.

Here we have to stress that it is difficult to compute one particular threshold fitting all three scores of each matching pair because of the formal language used in mathematics. Symbols, equations and eponyms may be contained in both titles and abstracts resulting in significant decrease of similarity scores. Therefore we decided to set a minimum threshold for each of the three matching parts separately to allow for a certain degree of variation. After manual accuracy check, an additional set of 4,018 arXiv papers could be identified. This way, we have increased the recall of the matching method by about 39% based on the results returned by Unpaywall. We use a set of arXiv papers containing the DOI information to validate the precision of the matching results, since the DOIs are supplied retroactively by the authors of arXiv papers after publication, or directly in the case of submitting an already published paper to arXiv, and considered to be reliable to be used as the evaluation dataset (Müller & Teschke, 2016). By matching DOIs of all the journal papers (45,200 out of 50,033 papers with DOIs in WoS) to those in the arXiv database via arXiv API, of which 3,564 arXiv papers are identified with their corresponding arXiv versions and by creating such an evaluation dataset, we performed the precision analysis and find it is estimated to be 96.70%.

In total, 10,310 journal papers with a version deposited in arXiv could be found by the time we collected them (October 2018), accounting for about 20.60% of the total 50,033 papers (see Table 1). Here we emphasize that the publication counts presented in the second column of Table 1 are not additive and cannot be summed up to the total because of the possibility of multiple sub-disciplines assignment. Furthermore, the proportion of arXiv papers is in line with what was reported by Larivière et al. (2014) for the publication year 2012. All the 10,310 arXiv papers are further subdivided into two types: *Preprint* - a paper submitted to arXiv before or at the same time as being published online in a journal; *Postprint* – a paper submitted to arXiv after being published online in a journal. In addition, in order to evaluate the effect of open access on preprints impact, we exclude all the open access (OA) papers published in *hybrid open-access journals* (subscription journals that allow open access publication as well) and *gold open-access journals* (journals listed on the Directory of Open Access Journals (DOAJ)), which may differ in terms of citations and the broader impact from non-OA papers and form a third category besides the toll-access papers with or without preprint versions. In order to guarantee the necessary statistical reliability, we have excluded journals publishing less than 10 preprints or 10 non-OA papers each, for the comparative analysis of the impact of articles in the two data sets whithin the same journals. This way, 7,669 preprints and 15,721 non-OA papers published in 220 journals could be selected for the preprint impact assessment. The distributions of the two data sets across sub-disciplines are shown in Table 1 (marked with the superscript-index '*' on the corresponding columns).

**Table 1.** The share of arXiv papers in Mathematics in 2013

| Sub-disciplines | Pubs | arXiv Papers | % arXiv | Preprints | Preprint Delay (Month) | JRs* | Pubs* | arXiv Papers* | Preprints* | Non-OA Papers* | Preprint Delay* (Month) |
|---|---|---|---|---|---|---|---|---|---|---|---|
| M1 | 26,588 | 7,788 | 29.29% | 7,371 | 17.21 | 147 | 15,206 | 5,919 | 5,665 | 8,584 | 17.10 |
| M2 | 26,005 | 4,307 | 16.56% | 4,021 | 14.21 | 91 | 12,376 | 3,131 | 2,901 | 8,495 | 13.81 |
| M3 | 9,302 | 459 | 4.93% | 376 | 12.23 | 15 | 2,008 | 309 | 276 | 1,540 | 10.89 |
| Total | 50,033 | 10,310 | 20.60% | 9,662 | 16.13 | 220 | 25,267 | 8,106 | 7,669 | 15,721 | 15.87 |

*Note: Sub-discipline codes*: M1–'Mathematics'; M2–'Mathematics Applied'; M3–'Mathematics Interdisciplinary Applications'; *Preprint Delay*: The average time elapsed from arXiv submission date to journal online date for preprints at the sub-discipline level; *JRs*\**: The selected 220 journals; *Pubs*\* \ *arXiv Papers*\* \ *Preprints*\* \ *Non-OA Papers*\**: The number of all publications(article & review) \ arXiv papers \ preprints \ non-OA papers published in the selected 220 journals; *Preprint Delay*\**:* the average *preprint delay* of preprints published in the selected 220 journals at the sub-discipline level.

## 3. Methodological aspects

In order to answer the above-mentioned research questions, we first quantify the publication delay of the preprints in the mathematics journals, and then build indicators of scholarly and social impact of preprints by using *multiple indicators*, including citations from WoS and Scopus (*Citations(WoS)*, *Citations(Scopus)*), Usage counts in WoS (*Usage(WoS)*), Mendeley readers (*Readers(Mendeley)*) and Twitter mentions (*Tweets*) from Altmetric.com. Preprints are compared with non-OA papers within the same journals. Finally, a regression analysis is conducted to assess the correlation between the different indicators.

*3.1. Publication process of preprints*

Two types of time designation are needed (as shown in Fig. 2): a. For preprint publication, the arXiv upload time, including arXiv submission time (i.e. the first time a manuscript being submitted to arXiv) and update time; b. For journal publication, the times of received, revised, accepted, online and print publication. arXiv provides the dates of upload for each e-print which can be accessed by using arXiv API. The online publication dates of all published arXiv papers can be retrieved by using Crossref API. By contrast, the received time and accepted time are only available on the publishers' websites, which are usually not allowed to be collected automatically. It is therefore impossible to collect this information for all papers used for this study. In order to fill the gap, we select the top 16 journals with the largest amount of arXiv papers (see Table 2), which have published at least 90 arXiv papers each as a proxy of the relationship between the submission/latest update time of the preprint version and received/accepted time of the journal version. The latter times are collected from the individual papers' entry links on the publishers' website.

**Table 2.** Sixteen journals with the largest amount of arXiv papers

| No. | Journal(sub-disciplines) | Publisher | Pubs. | arXiv Papers. | % arXiv | Preprints | Journal Delay** (Month) | Preprint Delay** (Month) |
|---|---|---|---|---|---|---|---|---|
| 1 | Advances in Mathematics *(M1)* | Elsevier | 299 | 233 | 77.93% | 231 | 26.19 | 18.24 |
| 2 | Journal of Algebra *(M1)* | Elsevier | 383 | 224 | 58.75% | 219 | 10.46 | 14.11 |
| 3 | Linear Algebra and Its Applications *(M1, M2)* | Elsevier | 635 | 194 | 30.55% | 177 | 23.14 | 9.41 |
| 4 | Journal of Mathematical Analysis and Applications *(M1, M2)* | Elsevier | 794 | 162 | 20.65% | 147 | 16.72 | 10.86 |
| 5 | Proceedings of the American Mathematical Society *(M1, M2)* | AMS* | 421 | 151 | 35.87% | 149 | 21.21 | 21.59 |
| 6 | International Mathematics Research Notices *(M1)* | Oxford Academic | 167 | 123 | 73.65% | 123 | 6.95 | 11.70 |
| 7 | Journal of Functional Analysis *(M1)* | Elsevier | 219 | 119 | 54.34% | 114 | 8.84 | 10.81 |
| 8 | Mathematische Zeitschrift *(M1)* | Springer | 188 | 118 | 62.77% | 115 | 10.79 | 14.99 |
| 9 | Journal of Differential Equations *(M1)* | Elsevier | 333 | 116 | 35.14% | 110 | 8.11 | 10.66 |
| 10 | Mathematische Annalen *(M1)* | Springer | 169 | 105 | 62.13% | 102 | 15.00 | 18.24 |
| 11 | Algebraic and Geometric Topology *(M1)* | MSP* | 117 | 101 | 86.32% | 99 | 10.08 | 15.75 |
| 12 | Transactions of the American Mathematical Society *(M1)* | AMS* | 229 | 99 | 43.23% | 99 | 23.23 | 25.28 |
| 13 | Journal of Pure and Applied Algebra *(M1, M2)* | Elsevier | 173 | 94 | 54.34% | 91 | 10.33 | 14.33 |
| 14 | Journal of Geometry and Physics *(M1)* | Elsevier | 138 | 93 | 67.39% | 90 | 6.89 | 11.79 |
| 15 | Journal of Combinatorial Theory Series A *(M1)* | Elsevier | 137 | 90 | 65.69% | 87 | 10.57 | 12.39 |
| 16 | Journal of Number Theory *(M1)* | Elsevier | 242 | 90 | 37.19% | 82 | 13.24 | 15.55 |
| | Total | | 4,644 | 2,112 | 45.48% | 2,035 | 14.96 | 14.74 |

*Note:* *AMS: American Mathematical Society; MSP: Mathematical Sciences Publishers;* **Journal Delay:* The average value of the time elapsed from the journal received date and journal online publication date of preprints at the journal level; *Preprints Delay:* The average value of the time elapsed from arXiv submission date and their journal online publication date at the journal level.

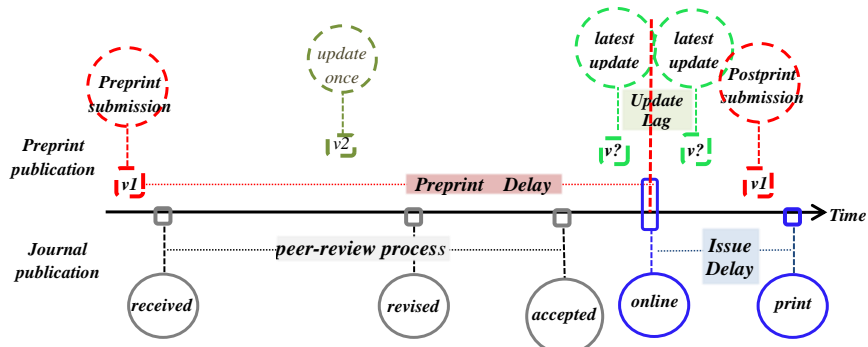

Note: Specific versions are referred to by adding the version number, i.e., v1, v2….
**Fig. 2.** The time designation of preprint and journal publications

*3.2. Scholarly and social impact indicators*

It is important to conduct deeper analysis on how preprint versions are cited by authors who published articles in journals, which reflects the attitude of a scientific community towards preprints. With respect to the second question, citations to all selected 7,669 preprint versions from WoS Core Collection database are identified in October 2018. Each arXiv paper is assigned with a unique arXiv identifier, which enables us to identify citations to it by using the specific structure of references in WoS. For example, a reference to the arxiv paper, 'arxiv: 0907.3987', contains the specific string 'ARXIV09033987'. Notice that the arXiv identifier scheme was changed since April 2007 and the old arXiv identifiers contain the subject class information followed by series of seven or eight digits, for example, references with the specific string 'ARXIVMATH0405285' are identified as citations to the arXiv paper, 'arXiv: math/0405285', coming from the Mathematics section in arXiv. In addition, the online publication dates of all the papers with the references to arXiv versions in WoS are retrieved by Crossref API and regarded as the citation time to the arXiv versions. By combing the arXiv preprints and their corresponding citing papers indexed in WoS, we are able to present a comprehensive picture for the scholars' communication behaviour in terms of citing preprints in formal publications and shed light on the characteristics of the aging patterns of preprint versions.

In order to answer the third question, a set of indicators ("*Multiple indicators*") is used to measure and assess the impact of preprints within scholarly communication and beyond (see Table 3). Different platforms provide different citation counts corresponding to their bibliographic coverage and are worthy being compared. In our analysis, two citation indicators, citations in WoS and in Scopus, denoted as "*Citations(WoS)*" and "*Citations(SC)*" respectively, are analysed to evaluate the citation impact of preprints. In addition, the three citation indicators – "*WoS_C13*", "*WoS_C15*" and "*WoS_C16-18*" – are designed to reflect the citation impact of preprints related to paper age: a first-year, short-term citation window ("*WoS_C13*", "*WoS_C15*") and a medium-term citation window based on three years ("*WoS_C16-18*") (cf. Moed, 2007; Wang et al. 2018). *Citations (WoS)* are collected from WoS Core Collection database and *Citations (SC)* are collected from Scopus. WoS usage counts (*Usage (WoS)*) are extracted from the WoS online version. Mendeley readers (*Readers(Mendeley)*) are collected from Mendeley API in the free software Webometric Analyst (cf. Webometric Analyst, 2019) by using DOIs. While for those papers without a DOI, we use their titles and first author last names in the API queries, and then by calculating the tf-idf similarity scores between the returned records and the retrieval item, we keep the returned record in Mendeley if it has at least 90% similarity with the WoS paper. For papers not found by using API or with "0" readers, we regard they don't have any readers in Mendeley and the number of *Readers (Mendeley)* is put zero. The number of *Tweets* originates from Altmetric.com by using Altmetric API. For preprints, arXiv IDs and DOIs are used to query Altmetric.com respectively and for non-OA papers, DOIs are used. When a publication is not covered in Altmetric.com or doesn't have any mentions in Twitter, the number of *Tweets* is put zero. In addition, in analogy to using the first citation as an indicator of response time by the scientific community (Schubert & Glänzel, 1986; Aman, 2014), the Altmetric '*first seen date*' (i.e., the date on which Altmetric.com captures the first event for a research output (Fang & Costas, 2018)) provided by Altmetric.com, is used as an indicator of the speed of discoverability of publications in social media. The *Altmetric attention delay*, measured by the time elapsed from the Altmetric '*first seen date*' to the arXiv submission date or journal online date, is compared among the three groups of altmetric data collected by arXiv IDs and DOIs of preprints and DOIs of non-OA papers respectively. Data for all indicators for preprints and non-OA papers are collected in October 2018.

**Table 3.** The definitions of relevant impact indicators in the study

| Impact | Indicators | Description |
|---|---|---|
| Citations | WoS_C13 / WoS_C15 / WoS_C16-18 / Citations (WoS) | The number of citations the document received from WoS until Dec. 2013 / until Dec. 2015 / from Jan. 2016 to Oct. 2018 / until Oct. 2018 |
| | Citations (SC) | The number of citations the document received from Scopus until Oct. 2018 |
| Usage | Usage (WoS) | The number of times the full text of a record has been accessed or a record has been saved on WoS since 1$^{st}$ February 2013 |
| Capture | Readers (Mendeley) | The number of people who have added the document to their private libraries on Mendeley |
| Social media attention | Tweets (a-DOI) | The number of Twitter users who have tweeted (or re-tweeted) the journal version of an arXiv paper. |
| | Tweets (a-ID) | The number of Twitter users who have tweeted (or re-tweeted) the arXiv version of an arXiv paper. |
| | Tweets | *For arXiv papers,* it is the total number of Tweets for publications with at least one *Tweets (a-DOI)* or *Tweets(a- ID)*.<br>*For non-arXiv papers*, it is the total number of Twitter users who have tweeted (or re-tweeted) a publication. |

*3.3 Impact Differential Ratio (IDR)*

In order to quantify and measure the impact differential of preprints versus non-OA papers, we use the optimised function *Impact Differential Ratio* (IDR), which is based on the function "*arXiv Citation Impact Differential (CID)*" proposed by Moed (2007), which had been optimised and has already been applied to citation analysis of preprints in Library & Information Science (LIS) by Wang et al. (2018). To guarantee the reliability of comparison, it is conducted between preprints and non-OA papers published in the same journals (cf. Harnad & Brody, 2004; Gargouri et al., 2010). We compute the values of IDR for each individual journal $j$ ($IDR_j$). Then we calculate the mean values for all 220 journals ($\overline{IDR}$). Hence, we obtain the final formulas as:

$$IDR_j = 200 \times \frac{Pre_j - Noa_j}{Pre_j + Noa_j} \; ; \; \overline{IDR} = \frac{1}{n}\sum_{j=1}^{n} IDR_j$$

Where $n$ is the number of journals; $Pre_j$ denotes the mean *Citations*, *Usage(WoS)*, *Readers (Mendeley)* or *Tweets (a-ID) / Tweets (a-DOI)* of preprints in journal $j$; and $Noa_j$ denotes the mean *Citations*, *Usage (WoS), Readers (Mendeley)* or *Tweets* of non-OA papers in journal $j$.

As indicated by Moed (2007), compared with OA versus non-OA IR defined as $IR = 100 \times \frac{Cpp_a}{Cpp_{na}}$ by Harnad and Brody (2004), where CPP denoted the number of received citations per article, and the indexes *a* and *na* denoted whether the cited paper was deposited in arXiv or not, the biggest advantage of *IDR* is more suited for our purposes due to its insensitivity to "small values" in the dominator, which may distort the effect to be measured. In particular, the preprints versus non-OA *IDR* is insensitive to this effect since $Noa_j \ll 1$ would simply result in *IDR* values close to 200 as the ratio on the right-hand side is about 1. The ratio $\frac{Pre_j}{Noa_j}$ would otherwise become extremely large, independently of the actual value of $Pre_j$, provided $Pre_j > 0$. Furthermore, using the same formula enables us to compare the results with the citation analysis of arXiv preprints in Condensed Matter (Moed, 2007) and in Library and Information Science (LIS) (Wang et al. 2018).

*3.4 Linear regression analysis*

In order to investigate the influence of a paper's preprint-deposited status on the impact differential between preprints and non-OA papers by controlling additional factors, regression analysis is conducted on the citation, usage, capture and social media attention indicators with a set of independent variables related to the authorship and article itself, all of which are known to associate with a paper's citations and altmetric impact. The variables include the 'preprint-deposited' status of journal paper which is coded as a binary variable, with a value of "1" for papers having a preprint version deposited in arXiv, and "0" for those without, the Journal Impact Factors (JIF) of the publication year (the year of 2015 in this case), which is available in the Journal Citation Reports (JCR) published by Clarivate Analytics, the number of authors (1, 2, 3, 4, 5+ authors), references, pages of a paper and its article type (review, article), sub-disciplines (MP, MA, MIA). It is noted that the number of authors of each paper is modelled separately for 1, 2, 3, 4, and 5+ authors following the similar approach employed by Thelwall & Sud (2020), since its distribution is far from normal even by using a log formula and more than 99% papers having no more than 5 authors of either the preprints or the non-OA papers.

In terms of authorship-related variables, the influence of country and academic age of the first and last author of each paper is tested in the regression analysis. The country of each author is extracted from the author addresses information in "C1" field of our WoS dataset, and then coded with a value of "1" or "0" based on the author's affiliation located in USA or not, since in our datasets, the proportion of USA-based authors in preprints is much higher than that in non-OA papers (23.5% vs. 13.5%) and it is already well-known to be positively correlated with the citation counts (Fraser et al., 2020; Fu & Hughey, 2019; Gargouri et al., 2010; Davis, Lewenstein, Simon, Booth, & Connolly, 2008). The academic age is usually defined as the time between the year of the first formal publication of an author and the publication year of the paper in question (Nane, Larivière, & Costas, 2017). In this paper, the academic age of individual authors is calculated as follows:

$$\text{Academic age} = 2015 - SY + 1$$

Where, SY is the publication year of the first formal publication of the author.

The most challenging to identify the year of an author's first publication is the author name disambiguation, especially for authors from Asian countries such as China (Wu & Ding, 2013; Han et al., 2017). To deal with the problem, Scopus Author Identifier (Scopus author ID), a unique number assigned automatically to each author in Scopus to group together all of the documents written by that author, are used to retrieve authors' publication histories in Scopus. The recall and precision of the Scopus author ID has been verified to be high by multiple researchers (Kawashima & Tomizawa, 2015; Moed, Aisati, & Plume, 2013) and has been successfully used in order to identify the first recorded publication in Scopus for authors of bioRxiv-deposited papers (Fraser et al., 2020) and identify authors of large publishing consortia (Thelwall, 2020). In consider that our datasets have a high coverage in Scopus (95.72% arXiv papers and 86.28% non-OA papers are also indexed in Scopus) and the well-performance of results of the author name disambiguation method used by Scopus, the Scopus author ID is prioritized to be used as a trade-off between data availability and processing times (Tekles & Bornmann, 2019). Table 4 summarizes the independent variables investigated in this paper.

Two regression models, the linear regression model using log-transformed citations, usage and altmetric data and the negative binomial regression model using raw data, have been suggested to be suitable for analysing the impact metrics, which typically have highly skewed distributions (Fraser et al., 2020; Ajiferuke & Famoye, 2015; Thelwall & Wilson, 2014). To determine which method is better, we accessed the relative goodness-of-fit for each of the two regression models via the Akaike Information Criterion (AIC; Akaike, 1974). The lower the AIC values, the better the goodness-of-fit. We first conducted a reduced regression model to test the influence of 'preprint-deposited' status in the absence of the other 15 independent variables in Table 4, and then, a full regression model including all variables. For all the models, the lower AIC values were reported by using the linear regression method. In addition, the log-transformation can prevent special articles with extreme performance in a specific impact metric from dominating the results to some extent (Thelwall & Sud, 2020). Thus we only report the results of linear regression analysis based on the reduced and full regression models, which are defined as follows:

$$I_{reduced\ model} = \alpha + \beta_1\ P$$

$$I_{full\ model} = \alpha + \beta_1\ P + \beta_2 JIF + \beta_3 R + \beta_4 Fage + \beta_5 Lage + \beta_6 Fus + \beta_7 Lus + \beta_8 A_2 + \beta_9 A_3 + \beta_{10} A_4 + \beta_{11} A_5 + \beta_{12} Ref\_n + \beta_{13} Page\_n + \beta_{14} MP + \beta_{15} MA + \beta_{16} MIA$$

Where $I$ denotes the log-transformed value of each of the multiple impact indicators (i.e., *Citations (WoS)*, *Citations (SC)*, *Usage (WoS)*, *Readers (Mendeley)*, *Tweets (a-DOI)*, *Tweets (a-ID)*) used in this paper. For the $I_{reduced\ model}$ only the binary independent variable 'Preprint-deposited (*P*, 1 / 0)' is included. For the $I_{full\ model}$, all the 16 independent variables listed in Table 4 are included. $A_i = 1$ if the paper has i authors (or $\geq i$ if $i = 5$, if all $A_i$ are 0 then the paper has one author). $\beta_i$ denotes the corresponding regression coefficient. In addition, the independent variables in all regression models have Variance Inflation Factors (VIF) below 5 (see Table 4), indicating acceptable levels of multicollinearity.

**Table 4.** Summary table of descriptive statistic for independent variables and their VIF in the linear regression model

| Independent Variables | Median | | Mean | | VIF |
|---|---|---|---|---|---|
| | Preprint | Non-OA | Preprint | Non-OA | |
| Preprint-deposited (*P*, 1 / 0) | | | | | 1.139 |
| Journal Impact factor (*JIF*) | 0.762 | 0.877 | 0.934 | 1.030 | 1.604 |
| Review article (*R*, 1 / 0) | | | 0.0031 | 0.0016 | 1.004 |
| First author academic age (*Fage*)* | 10 | 10 | 13.30 | 13.64 | 1.104 |
| Last author academic age (*Lage*)* | 11 | 13 | 14.23 | 15.78 | 1.118 |
| First author is from USA (*Fus*, 1 / 0) | | | 0.23 | 0.13 | 1.857 |
| Last author is from USA (*Lus*, 1 / 0) | | | 0.24 | 0.14 | 1.865 |
| Two Authors (*A₂*, 1 / 0) | | | 0.39 | 0.38 | 1.440 |
| Three Authors (*A₃*, 1 / 0) | | | 0.20 | 0.23 | 1.420 |
| Four Authors (*A₄*, 1 / 0) | | | 0.05 | 0.07 | 1.199 |
| Five Authors or more (*A₅*, 1 / 0) | | | 0.02 | 0.03 | 1.096 |
| Number of References (*Ref_n*)* | 22 | 20 | 24.60 | 23.45 | 1.440 |
| Number of Pages (*Page_n*)* | 20 | 14 | 22.29 | 16.43 | 1.330 |
| MP (1 / 0) | | | 0.74 | 0.57 | 1.985 |
| MA (1 / 0) | | | 0.38 | 0.60 | 1.496 |
| MIA (1 / 0) | | | 0.04 | 0.11 | 1.297 |

Note: * Log transformed

## 4. Results

### *4.1. Publication delay of preprints*

In this study, the publication delay of arXiv papers (i.e. preprints and postprints) is defined as the time lag between the date of being first submitted to arXiv and the date of being published online in a journal. The results are shown in Fig. 3, in which the time lag is interpolated and converted to months and numbers are negative or positive according as the paper was submitted to arXiv earlier or later than its online date for the journal publication. The mean and median publication delay of preprints (see Fig. 3, left) is -16.13 months and -13 months, respectively. The corresponding values for postprints (see Fig. 3, right) are 14.3 months and 8 months, respectively. Almost all arXiv papers (about 94.0%) are submitted before or simultaneously with the journal (online) publication. Most of the authors submit the first version of arXiv papers seven or eight months prior to the journal online publication (Fig. 3, left). Furthermore, more than two thirds of preprints are posted within one year and 80.66% are posted within two years before being published online in journals.

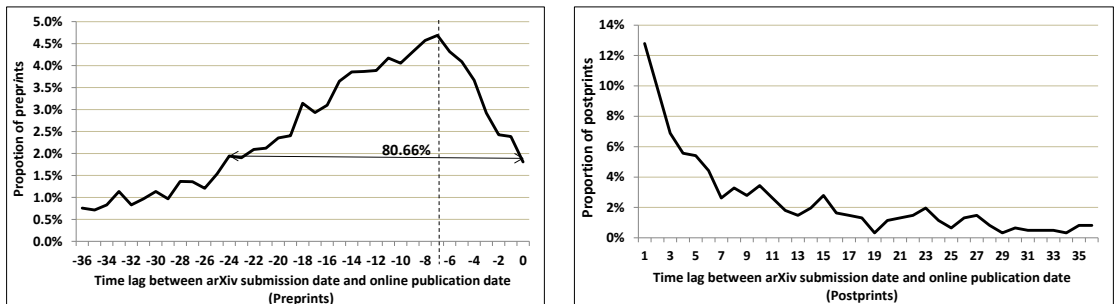

**Fig. 3.** Publication delay of preprints (left) and postprints (right) (arXiv papers deposited more than 36 months before or after online publication are omitted)

By contrast, only less than 6% of the total 10,310 arXiv papers in Mathematics are postprints (see Fig. 3, right), and they are always posted on arXiv immediately after being published in journals. For all arXiv papers (i.e. preprints and postprints), an average of 5.1 months is still needed to be finally published in printed form, denoted as '*Issue Delay*' in Fig. 2. In total, the average time duration from arXiv submission date to journal print publication date of preprints in Mathematics is longer than 21 months, which is much longer than Physics (6 months), Computer Science (12 months) (Larivière et al., 2014) and Biology (5.5 months) (Abdill & Blekhman, 2019). It is worth mentioning that the metric on the average *preprint delay* varies markedly among the three sub-disciplines (see Table 1): compared with the other two sub-disciplines, 'Mathematics' (M1) has the longest *preprint delay* (17.21 months), followed by M2 (14.21 months), while M3 has the shortest (12.23 months). To summary, the preprint publishing benefits researchers in Mathematics from bypassing the publication delay led by the journal peer-review process, and enabling researchers to communicate research findings in an easy and fast way.

We further investigate the authors' preprint-publishing behaviour in the top 16 journals listed in Table 2 by quantifying the time interval between a preprint submission or updated date on arXiv and dates in its peer-review process. Fig. 4a shows that most authors submit preprints to arXiv almost at the same time as submitting them to journals. 1,363 out of the 2,030 preprints (excluding the five preprints missing the received date on the publisher website) published in the 16 journals are submitted to arXiv before being received by journals, among which 488 (35.8%) preprints are submitted to arXiv and journals on the same day. In contrast, for preprints that have arXiv submission time later than journal received time, a majority (42.8%) of them have the "submission-received" duration less than one month, accounting for 9.3% of the total 2,030 preprints. The results indicate that most authors prefer to post a preprint version of the manuscript on arXiv before or very close to submitting it to a peer-review journal, with the desire to share their new research outputs as soon as possible and receive useful feedbacks from wider readers.

Compared with the trend of the publication delay of preprints in Fig. 3, the average value of *preprint delay* of the 16 journals, all of which are actually assigned to M1 ('*Mathematics*') of the WoS Subject Categories, is shorter (448.6 days, i.e., 14.74 months) (Fig. 4b). While it is important to notice that the *preprint delay* shows considerable variation across the journals given in Table 2, i.e., four journals with *preprint delay* longer than 18 months, while four other journals with preprint delay less than 11 months. As shown in Fig. 4b, most of the preprints are submitted to arXiv between 5 and 10 months prior to their journal online publication time. More than half (53.7%) of the total 2,035 preprints are posted on arXiv at least 12 months earlier than being published online in journals, and 78.9% are more than 6 months earlier. The trend of the latest update time of arXiv papers (see Fig. 4c and 4d) provides additional information on the functions that preprints have. arXiv authors could promptly update new versions to make changes such as correction, addition or extension, with a new date stamp generated (cf. ArXiv, 2020). The changes indicated by authors appear in '*Comments:*' field of the new submission, which could help readers know why a paper is replaced. 1,082 preprints (i.e., 53.2% of all 2,035 preprints) are updated at least once after the first submission to arXiv, and among them, 60.0% are updated once, 26.5% are updated twice and 8.2% are updated three times. While all of the 1,082 preprints have a journal online publication time, only 732 of them could be found with accepted or revised time information. Therefore, Fig.4c is the result of the 732 preprints, which shows that more than 57.1% papers are replaced by new versions after being accepted by journals or after the time of the last revision of the manuscript, most of the preprints (26.40%) are updated within one month, which is usually very close to its accepted time, and only 9.8% of the preprints have the latest updated time six months later than their accepted/revised time. Fig. 4d shows that 35.0% of all 1,082 preprints are updated within three months before being available on the journal websites, 9.2% are immediately updated within one month after, and less than 5.4% are updated 12 months later.

Previous studies suggested that the publishers' copyright and self-archiving policies would significantly influence authors' choices on self-archiving (Pinfield, 2004). The top 16 journals listed in Table 2 are published by five publishers, all of which have definite open self-archiving policies: Elsevier[1], AMS[2] and MSP[3] allow authors to immediately update the preprints with the accepted manuscript after peer review via arXiv.org; Oxford Academic[4] and Springer[5] require 12 months embargo periods after the first publication in journal. Because fourteen out of the top 16 journals are published by Elsevier, AMS or MSP, our results shown in Fig. 4c and Fig. 4d indicate that authors update the preprint version *after* journal acceptance or publication, that is, probably updated with the peer-reviewed and accepted manuscript, and readers who do not have access to the toll-access journal version are able to read the final version via the preprint repository. All of the above results suggest that preprint publishing speeds up scholarly communication through the early and fast dissemination, immediately update of new versions and removing the subscription or charge barriers to access. Compared with non-OA papers published in the same journals, preprints have the potential benefits of wider readership and earlier accumulation of attention or even citations contributed by the 'open-access' and 'early-view' effects. In the following, we would provide new and deeper insights to both scholarly and broader impact of preprints by using bibliometric and altmetric analysis.

---

[1] https://www.elsevier.com/about/policies/sharing
[2] https://www.ams.org/publications/journals/open-access
[3] https://msp.org/publications/policies/
[4] https://academic.oup.com/journals/pages/access_purchase/rights_and_permissions/self_archiving_policy_b
[5] https://www.springer.com/gb/open-access/publication-policies/self-archiving-policy

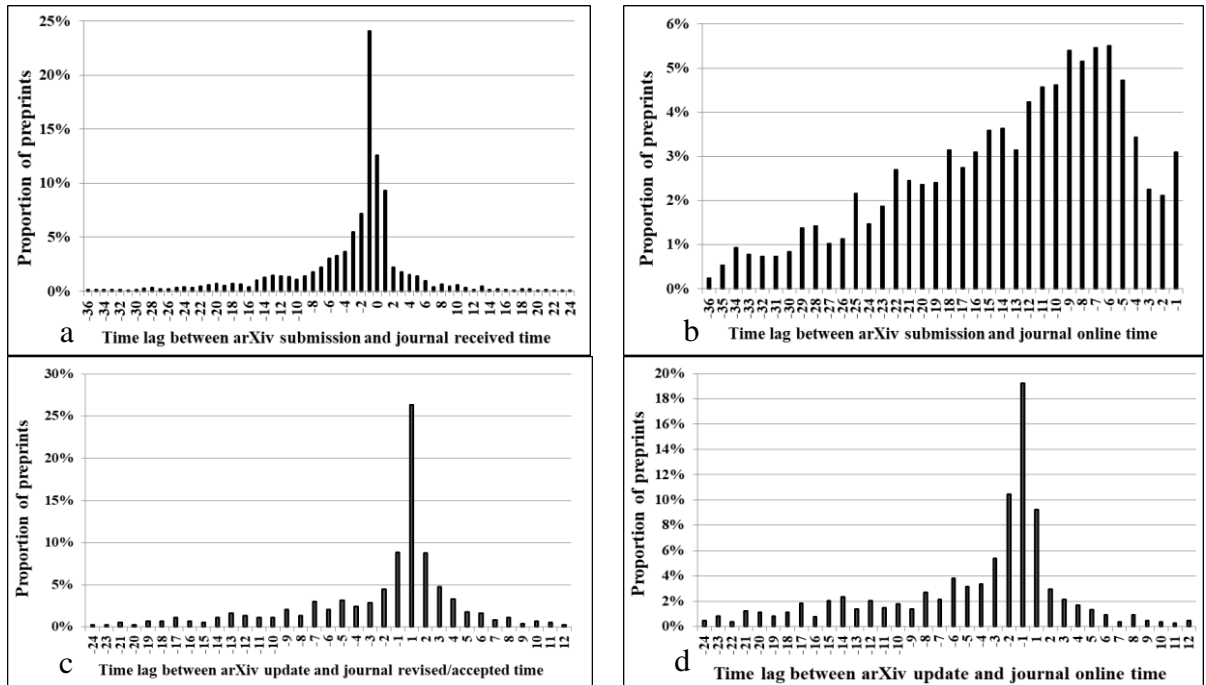

**Fig. 4.** Time lag between arXiv submission / latest update time and journal received / accepted / online publication time in the top 16 journals with the largest amount of arXiv papers

*4.2. Citation characteristics and aging patterns of preprints*

Studying how preprints are cited in academic papers is helpful to learn about not only the scholars' communication behaviour in terms of citing preprints in peer-reviewed journal publications, but also the role of preprints in the peer-reviewed journal publishing system. Mathematics is one of the fields that heavily relies on communication through preprints (Li, Thelwall, & Kousha, 2015) and where preprints are always treated as the same as the published versions (Davis & Fromerth, 2007). Evidence is given that, papers deposited in arXiv are increasingly cited by web sources indexed by Google Scholar (Noruzi, 2016) and scholarly documents indexed in Scopus (Li et al., 2015). arXiv papers are assigned with a unique arXiv identifier, which allows unambiguous identification and assignment of citations received by arXiv papers. These citations can, for example, be collected from WoS by using the arXiv identifiers.

For all 7,669 arXiv preprints in our study, 1,495 preprints (19.5%) are cited by 2,873 WoS papers until October 2018, and the total number of citations amounts to 3,183. However, the citation distribution of the preprint versions on arXiv is very skew (cf. Fig. 5), more than 61.3% of all cited papers have been cited only once, and the number of citations received by the top 16 cited papers with at least 10 citations each amounts to 19.1% of the total. The most cited preprint (arXiv: 0907.3987; doi: 10.1016/j.aim.2012.09.027) has received 293 citations from WoS indexed documents.

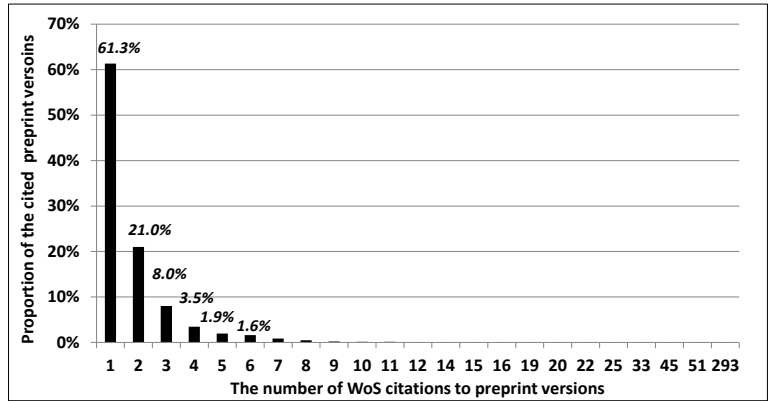

**Fig. 5.** The distribution of WoS citations to the preprint versions in arXiv

The aging distributions of citations to the arXiv versions as reflected by WoS citations (see Fig. 6) show that about 50% of all these citations are received within 24 months after the first arXiv submission time, above all, between the 12[th] and the 24[th] months, accounting for more than 40% of all citations (cf. Fig. 6a). Such trend can also be observed in Fig. 6c. Most of the 1,495 preprints receive their first citation in WoS during the 10[th] to 20[th] months (the percentage of these preprints ranging from 2.8% to 4.2%) after being disposed on arXiv, while beyond 20 months, the probability to get cited decreases dramatically. The reason for such an aging pattern can be explained by Fig. 6b. There is a significant decline of the average citation rate of preprints after journal online publication (Fig. 6b), which confirms the observation by Wang et al. (2018) that authors prefer to cite the journal version rather than the preprint version when both are available. In addition, 71.8% of all 1,495 cited arXiv versions are actually cited before publication time, accounting for 62.5% of all 3,183 citations, and the longest time lag between cited time and journal online publication time is 67 months. However, the two versions (i.e. arXiv preprint version and journal version) of the same papers have deviating readership profiles. For the preprint versions, 27.5% of the total citations in WoS come from papers assigned to Physics, while for their journal versions, the corresponding percentage amounts to 12.5%. This means that depositing papers on preprint server reaches a more heterogeneous readership in terms of their research profiles.

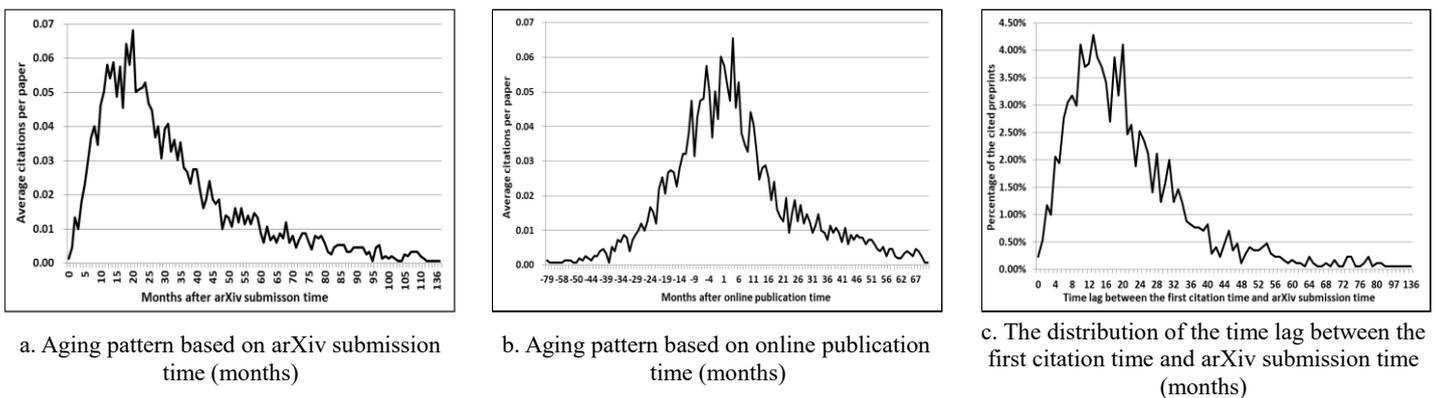

| a. Aging pattern based on arXiv submission time (months) | b. Aging pattern based on online publication time (months) | c. The distribution of the time lag between the first citation time and arXiv submission time (months) |

**Fig. 6.** Aging patterns of citations from WoS to the preprint versions

We further investigate the distribution of citations over time to the journal versions of preprints and non-OA papers. Here we emphasize that the citations to preprint versions are not taken into account, since we aim at analysing the effect of preprints on the citation impact of their final journal versions. As shown in Fig. 7a, the citing process starts earlier than the print publication year (i.e., the year of 2013 in our case) because of the time gap between the online and the print version of the journal papers (i.e. the 'Issue Delay' in Fig. 2). Although citation rates in the three sub-disciplines (M1–M3) largely differ, each sub-discipline shows the same pattern, and so does the complete field defined as the combination of the three sub-disciplines. The mean citation rate achieved by preprints is higher than that by non-OA papers in each year.

Furthermore, if we group preprints by arXiv submission year, the citation aging patterns of the journal versions in each group become different. Fig. 7b presents the results for preprints in the sub-discipline M1, which is actually the largest sub-discipline in Mathematics in terms of the number of preprints. The outcomes for the other two sub-disciplines are quite similar to M1. We have to note that only preprints posted to arXiv in the period from 2009 to 2013 are taken into account, since the number of preprints in other years proved too small to allow any reliable statistical analysis. The earlier a preprint is submitted to arXiv, the sooner and more citations are received by its journal version, especially in the first two years after journal publication. While citations to papers in the group "arXiv:2013" keep increasing within the 5-year citation windows, its average citation rate is the slowest compared to the other groups in each year. The citation patterns of preprints in Mathematics on a large scale confirm the observations by Gentil-Beccot, Mele, & Brooks, (2010) in the context of the citation patterns of arXiv papers in High-energy Physics, in which he emphasised that it was the wider and earlier dissemination of arXiv papers that contributed most to the citation advantage of the final journal articles.

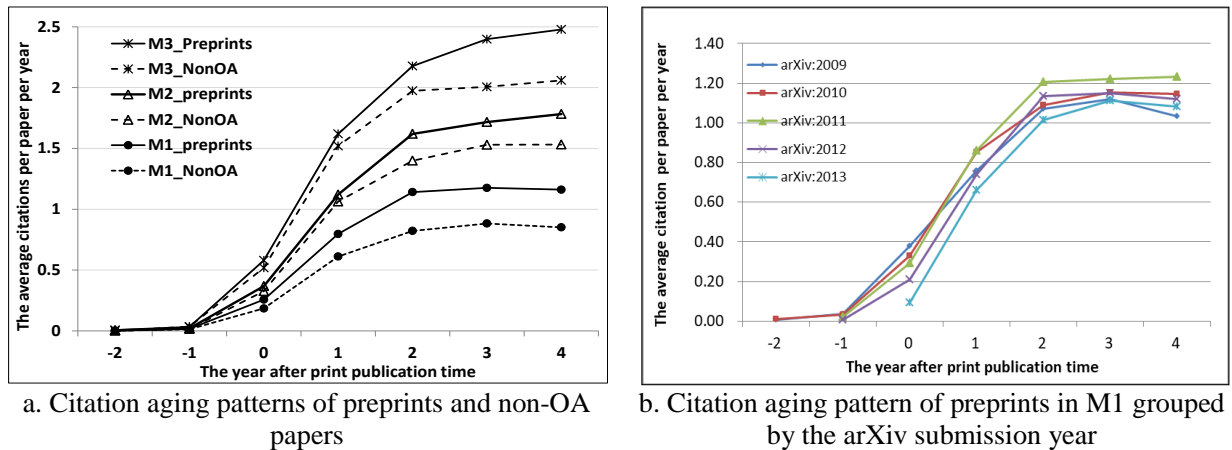

a. Citation aging patterns of preprints and non-OA papers

b. Citation aging pattern of preprints in M1 grouped by the arXiv submission year

**Fig. 7.** The citation aging patterns of preprints

It is also interesting to investigate what's the relationship between the citations to preprint versions and the citations to their (later) journal versions (denoted as "*Preprint-CT*" and "*WoS-CT*" respectively for short and both are the total citations received from WoS Core Collection database by the collecting time October 2018). Considering that both the citation distributions of preprint versions and journal versions are highly skewed, Spearman Correlation analysis is used for the total 7,669 preprints data set in this paper. The correlation coefficient between the two indicators is $0.24^{**}$ ($\rho < 0.01$), which is regarded to be a low but statistically significant positive relationship. We further deepen the result by conducting the regression analysis of "*Preprint-CT*" versus "*WoS-CT*" conditional expectation, which has been successfully applied to the correlation between author self-citations and foreign citations (Glänzel & Thijs, 2004), citation impact and download statistics (Glänzel & Heeffer, 2014) and usage counts in WoS (Chi & Glänzel, 2018). As shown in Fig. 8, the mean WoS citation rates of the journal versions have been calculated under the condition that the corresponding preprint versions have a given number of WoS citations. We need to stress that in order to avoid the small sample size distorting the results (in this case, resulting in huge fluctuations at the high end of the citation counts scale), we truncated at a point beyond which the sample size drops below ten. By using such a method, the relationship between "*Preprint-CT*" and "*WoS-CT*" is more significant ($R^2 = 0.66$), and the slope of the regression function reveals there is a "translation factor" between them. Such a method is also applied to reveal the relationships between citations and the other two indicators, usage counts in WoS (*Usage(WoS)*) and Mendeley readers (*Readers(Mendeley)*), in the following content.

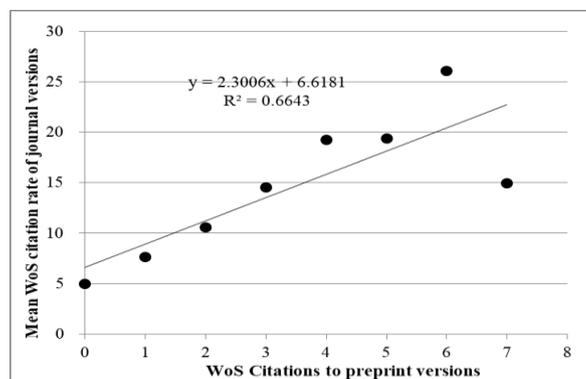

**Fig. 8.** The plot of conditional mean WoS citation rate of journal versions as a function of WoS citations to preprint versions

*4.3. The comparison of citations, usage, capture and social media attention indicators of preprints*

In order to provide an overview of our dataset, we first show the mean and median values of each impact indicator of preprints and non-OA papers in the three sub-disciplines of Mathematics in Table 5. In Mathematics, citations accumulated in 5-year windows are higher than *Usage(WoS)*, *Readers(Mendeley)* and *Tweets*, except for M3, where a higher *Usage(WoS)* is observed. The *Citations(SC)* are somewhat higher than *Citations(WoS)* for both data sets, which is a result of the broader coverage of Scopus. *Readers(Mendeley)* of preprints are always more frequent than their *Usage(WoS)*, while the *Usage(WoS)* of non-OA always exceeds those of preprints. The presence and density (i.e. the average number of *Tweets* per publication) of Tweets among either of the two "document types" (i.e. preprints and non-OA papers) is much lower than that of *Readers(Mendeley)*, but there is a greater advantage for preprints in attracting attention on social media platforms, especially for the preprint versions on arXiv. More than 22.7% of preprints are tweeted with arXiv IDs, while only 4.70% of them are tweeted with DOIs. But both are higher than that of non-OA papers, with only 2.37% being mentioned on Twitter.

Preprint versions are not only more frequent in social media, but also tracked by Altmetric.com much quicker than their final journal versions and non-OA papers. As the cumulative distribution trend of *Altmetric attention delay* presented in Fig. 9, the majority (75.50%) of arXiv IDs have received any altmetric score within five days after being submitted to arXiv, in contrast, the proportion is only 44.49% and 29.98% among the journal versions of preprints and non-OA papers tracked by DOIs respectively. It is noted that the 507 non-OA papers tracked by Altmetric.com are published in 107 journals, and in fact, 121 (23.87%) are published in *International Journal of Quantum Chemistry* (denoted as '*IJQC*' for short). In addition, for the 118 (23.27%) non-OA papers with no more than '1' day of the *Altmetric attention delay* (see Fig. 9), 99 papers (83.89%) are from *IJOC*. By a further look at the sources contributing to the first Altmetric attention score of the 99 papers, we find that the majority of the Altmetric events come from Twitter, because the journal has an official Twitter account and tweets with links to the online version with a DOI as soon as the paper is available on the journal website. Another interesting result is that for the 504 DOIs of preprints covered in Altmetrics.com, 199 (39.48%) have the Altmetric '*first seen date*' earlier

than the journal online date (see Fig. 9, the negative value '<0' on the X-axis), due to their preprint versions making them to be available in social media and tracked by Altmetric.com before the online publication on the journal website, accounting for 70.07% of all the 284 preprints having Altmetric IDs with both DOIs and arXiv IDs. The results indicate that the patterns of knowledge dissemination in social media differ from those in traditional peer-review journals.

Although the citation trends presented in Fig. 6 indicates that that authors prefer to cite the journal version rather than the preprint version when both are available, the results presented in Table 5 and Fig. 9 indicate the preprints versions with arXiv IDs are more likely to be mentioned in social media than the "more official" journals versions. These differences in the reasons for the 'usage' of preprints can at least partially explained by user types of the different communication platforms. Traditional subscription databases target scientific researchers working in organizations with availability of the databases. While the audience in Twitter are much more heterogeneous, including more interested general public outside academia who may lack access to the final journal versions, therefore, publications may gain more potential readers if a link to a green open-access preprint version (i.e. arXiv version) is provided. In addition, posting the preprints on social media platforms can also facilitate the interactions between authors and readers, in particular, experts and the general public (Kim, 2010). Readers can post comments in social media or contact the researchers directly. The feedbacks may be useful for authors to revise the manuscripts before submitting to a journal or in addition to the comments from reviewers in traditional peer review.

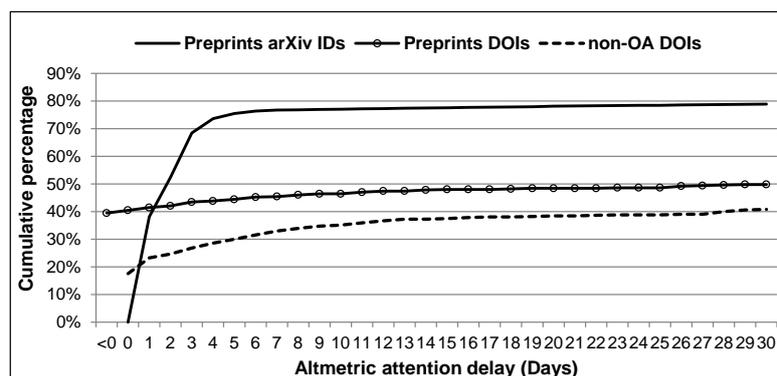

**Fig. 9.** The cumulative distribution of *Altmetric attention delay* of preprints vs. non-OA papers tracked by Altmetric.com

All values of the IDR for different impact indicators of preprints are shown in Fig. 10. The average values of IDR for all citation indicators, *Readers(Mendeley)* and Twitter metrics (*Tweets(a-ID)* and *Tweets(a-DOI)*) for preprints across journals are significantly larger than zero, and for the *Usage(WoS)*, we observe the opposite trend. Based on a correlated-sample *t*-test for all 220 selected journals we obtain ρ values below 0.01. Preprints received statistically significant more citations than non-OA papers in each of the three fixed citation widows and across the three sub-disciplines of Mathematics. The IDR values of *WoS_C13* are higher than those of *WoS_C15* and *WoS_C16-18* in M1 and M2, which might be a consequence of the "early-view" effect of preprints: preprints have more time to attract readers and receive citations due to they being free availability via arXiv usually much earlier than via journals.

**Table 5.** Mean and median values of the multiple indicators in the three sub-disciplines of preprints and non-OA papers in Mathematics

| Sub-discipline | JRs | Type | Preprint delay | | Scholarly and social indicators (Mean / Median) | | | | | | | | | |
|---|---|---|---|---|---|---|---|---|---|---|---|---|---|---|
| | | | Mean | Median | *Citations (WoS)* | *Citations (SC)* | *Usage(WoS)* | *Readers (Mendeley)* | Cov. % | *Tweets* | Cov.% | *Tweets (a-ID)* | *Tweets (a-DOI)* | Cov.% (a-ID \| a-DOI) |
| M1 | 147 | Preprints | 17.10 | 14 | 5.24/3 | 5.46/3 | 1.95/1 | 2.76/2 | 73.56% | 0.29/0 | 22.12% | 0.27/0 | 0.05/0 | 21.34%\|3.79% |
| | | Non-OA | | | 3.86/2 | 4.01/2 | 3.02/2 | 1.63/1 | 55.53% | 0.01/0 | 0.82% | n/a | n/a | n/a |
| M2 | 91 | Preprints | 13.81 | 11 | 7.67/4 | 8.54/5 | 4.46/3 | 5.06/3 | 79.73% | 0.38/0 | 24.41% | 0.36/0 | 0.11/0 | 23.09%\|5.89% |
| | | Non-OA | | | 7.01/4 | 7.88/4 | 7.69/4 | 4.08/2 | 63.38% | 0.025/0 | 1.68% | n/a | n/a | n/a |
| M3 | 15 | Preprints | 10.89 | 8 | 10.59/6 | 11.38/7 | 10.90/7.5 | 10.66/6 | 85.51% | 1.18/0 | 38.04% | 1.14/0 | 0.67/0 | 35.51%\|13.04% |
| | | Non-OA | | | 9.34/5 | 10.75/5 | 19.46/14 | 7.87/4 | 86.04% | 0.12/0 | 9.29% | n/a | n/a | n/a |

Note: Sub-discipline code: M1–'Mathematics'; M2–'Mathematics Applied'; M3–'Mathematics Interdisciplinary Applications'.

**Table 6.** Spearman correlation coefficients between *Citations (WoS)* / *Citations (SC)*, *Usage (WoS)*, *Readers (Mendeley)* and *Tweets* [a]

| Code | *Citations (WoS)* vs. *Citations (SC)* | | *Citations (WoS/SC)* vs. *Usage (WoS)* | | *Citations (WoS/SC)* vs. *Readers (Mendeley)* | | *Citations (WoS/SC)* vs. *Tweets* | | *Usage (WoS)* vs. *Readers (Mendeley)* | | *Tweets* vs. *Usage (WoS)/ Readers (Mendeley)* | |
|---|---|---|---|---|---|---|---|---|---|---|---|---|
| | Pre | Non OA | Pre | Non OA | Pre | Non OA | Pre | Non OA | Pre | Non OA | Pre | Non OA |
| M1 | .95** | .93** | .21**/.22** | .27**/.28** | .24**/.26** | .24**/.27** | .063*/.05 | -.06/-.05 | .18** | .22** | .04/.07* | .06/.17 |
| M2 | .97** | .97** | .33**/.33** | .42**/.43** | .38**/.39** | .36**/.38** | .11**/.12** | .08/.08 | .33** | .37** | .24**/.21** | .19*/.03 |
| M3 | .97** | .96** | .28**/.30** | .40**/.40** | .34**/.35** | .40**/.39** | .18/.19 | .14/.14 | .52** | .40** | .30**/.35** | .18*/.09 |

Note: ** The values are significant at the level ρ<0.01; * The values are significant at the level ρ < 0.05 ; [a] Correlations between *Tweets* and the other four indicators (i.e., *Citations(WoS), Citations(SC), Usage(WoS)* and *Readers(Mendeley)*) are calculated for publications with at least one *Tweet*.

In M3 the citation advantage seems to increase with time. However, we found that there was just one non-OA paper that received one citation in 2012 and 2013, respectively, among the 10 preprints and 26 non-OA papers published in *Advances in Complex Systems* in 2013, when we checked the 15 journals in M3. The IDR of *WoS_C13* of the journal was consequently -200.0 and had a negative effect on the "performance" of *WoS_C13* in M3 due to the low number of journals in M3. The value of the IDR of *WoS_C13* in M3 would jump up to 21.8%, if the journal was excluded. However, the average IDR values of *WoS_C13*, *WoS_C15*, *WoS_C16-18* and *Citations (WoS)* based on all 220 journals are not significantly deviating from each other, but are all significantly above the zero baseline, indicating that the citation advantage of preprints versus non-OA papers is stable over the years.

Overall, the average IDR value of *Citations (WoS)* of all preprints in Mathematics is about 20% indicating that preprints have a significant citation advantage in WoS, which is slightly higher than in Scopus, although the latter one reports higher average number of citations per paper. Compared with the values of '*arXiv CID*' in Condensed Matter and LIS, which is 80% (Moed, 2007) and 95% (Wang et al., 2018) respectively, the value of IDR for the citations of preprints in Mathematics is relatively lower. In consideration of the lower mean citations and longer citation half-life to journal papers in Mathematics (Davis & Fromerth, 2007), it is expected that greater citation advantage of arXiv papers in Mathematics would be detected if longer observation period are given. The advantage of preprints becomes most visible in capture (i.e. *Readers (Mendeley)*) and social media attention (i.e. *Tweets (a-DOI) and Tweets (a-ID)*) impact, but not reflected by *Usage (WoS)*, which is much higher for non-OA papers than for papers having a preprint version deposited in arXiv.

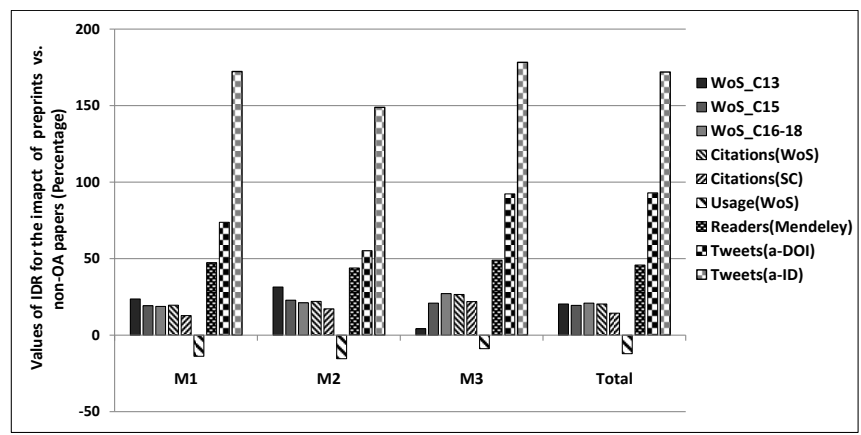

**Fig. 10.** The impact differential of preprints vs. non-OA papers based on multiple indicators

The significant differences in the IDR values of the scholarly and social impact of preprints measured by the citation, usage, capture and social media attention indicators suggest that compared with non-OA papers in Mathematics, preprints have a distinct citation advantage and enjoy a broader readership, not just from users of the WoS database. While articles with preprint versions deposited in arXiv achieve less usage counts than those without according to WoS, citation and readership advantage of preprints tell against any assumption that arXiv-deposited articles would be of lower "quality" or "value". A possible explanation for this phenomenon might be that gaining information through the WoS database might be especially relevant whenever no open-access information is available, which to some extent supports the explanation by Davis and Fromerth (2007) for the significant fewer full text downloads received by journal papers deposited in arXiv from the publisher's website.

*4.4 Regression analysis*

The above results show that there is significantly impact differential between the preprints and non-OA papers based on multiple indicators. However, we noted that there are some key differences in the various variables (see Table 4) that may influence the scholarly and social impact between the "two types" datasets (i.e., preprints and non-OA papers). To investigate the relationship between the citation, usage, capture and social media attention indicators and the 'preprint-deposited' status by controlling other influential factors (see Table 4), we conduct the linear regression analysis based on the reduced model (defined as '$I_{reduced\ model}$') and the full model (defined as '$I_{full\ model}$', see the 3.4 section of this paper) respectively, and the results are reported in Table 7. Results from the reduced regression model confirm the results shown in Table 5 and Fig. 10 that journal papers having preprint versions deposited in arXiv receive more citations, Mendeley readers and social attention in Twitter, but less usage counts in WoS than non-OA papers. Results from the full regression model indicate that the 'preprint-deposited' status remains an important independent predictor of *Citations (WoS / SC), Usage (WoS), Readers (Mendeley)* and *Tweets (a-DOI / a-ID)* even when controlling for the additional variables. In addition, it is noted that the *β* values of 'preprint-deposited' variable from the full regression model are almost the same as (or even slightly higher than) those from the reduced model for the multiple impact indicators, except *WoS (Usage)*, for which the *β* value of 'preprint-deposited' status is decreased (i.e., $β_{full\ model}$ = -0.218*** vs. $β_{reduced\ model}$ =-0.464***), but its influence remains statistically significant ($β_{full\ model}$ (95% CI) = -0.218*** (-0.241 - -0.196) ). Our results are consistent with other research on the citation and altmetric advantage of bioRxiv-deposited papers (Fraser et al., 2020; Fu & Hughey, 2019; Serghiou & Ioannidis, 2018), and suggest that previous preprint versions are associate with the citations, usage, capture and social media attention impact of their journal versions in Mathematics.

**Table 7.** Summary table of the results of the influence of the 'preprint-deposited' status for the outcome variables from reduced and full regression model

| Outcome Variables | $β_{reduced\ model}$ (95%CI) | Std. Error | $β_{full\ model}$ (95%CI) | Std. Error |
|---|---|---|---|---|
| *Citations (WoS)* | 0.137*** (0.11-0.164) | 0.014 | 0.138*** (0.113-0.163) | 0.013 |
| *Citations (SC),* | 0.134*** (0.106-0.162) | 0.014 | 0.149*** (0.123-0.174) | 0.013 |
| *Usage (WoS)* | -0.464*** (-0.491 - -0.437) | 0.014 | -0.218*** (-0.241 - -0.196) | 0.011 |
| *Readers (Mendeley)* | 0.155*** (0.13 - 0.18) | 0.013 | 0.249*** (0.225 - 0.272) | 0.012 |
| *Tweets (a-DOI)* | 0.024*** (0.019 - 0.028) | 0.002 | 0.033*** (0.028 - 0.038) | 0.003 |
| *Tweets (a-ID)* | 0.171*** (0.164 - 0.178) | 0.004 | 0.182*** (0.175 - 0.189) | 0.004 |

Note: *** $p < .001$; See Table A in the Appendix for the full results from full regression model including all independent variables.

In order to study the value of the new metrics in measuring and monitoring the impact of preprints, a regression analysis is conducted to examine the pairwise correlations among *Citations (WoS), Citations (SC), Usage(WoS), Readers(Mendeley)* and *Tweets*. The Spearman Correlations between the five indicators across three sub-disciplines for preprints and non-OA papers are presented in Table 6. The *Citations(WoS)* and *Citations(SC)* are highly correlated with each other and have low and medium correlations with *Usage(WoS)* and *Readers(Mendeley)* across three sub-disciplines. By contrast, the correlations between *Tweets* and *Citations (WoS/SC)* are much lower in both data sets, even though here we only consider those papers with at least one *tweet* to eliminate the bias caused by the high proportion of publications without any tweets being received by the time when we collected the data. Given the significant impact advantage of preprints in Tweets presented in Fig. 10, the very low correlations suggest that Twitter-based metrics have the potential to be used as altmetric indicators to capture and measure other interesting facets of impact of preprint publications, particularly the impact in the social public outside academia (Costas et al., 2015; Thelwall et al., 2013), which are hard to be detected by citation-based indicators and are important for understanding and assessing preprints impact. In addition, the correlation coefficients for M2 and M3 are close to each other, and both higher than those in M1. For all three sub-disciplines, the coefficients of "*Citations(WoS/SC)* vs. *Readers(Mendeley)*" are slightly higher than those of "*Citations(WoS/SC)* vs. *Usage(WoS)*" for preprints, while the relationship of the two pairs' coefficients are reverse for non-OA papers.

The conditional expectation regression analysis between *Citations (WoS)*, *Usage (WoS)* and *Readers (Mendeley)* for M1 (see Fig. 11) reveals another interesting aspect. This approach is of especial advantage if observations take similar values and form a cloud, usually at the origin of the coordinate system (cf. Glänzel and Thijs, 2004). Here *Citations (WoS)* are shown as pars pro toto because, on one hand, similar correlations can be found with *Citations (SC)*, and on the other hand, WoS covers the largest number of papers in our dataset. The different slopes of the regression lines for *Readers (Mendeley)* and *Usage (WoS)* vs. *Citations(WoS)* and for the two "document types" can be considered a kind of specific "translation factors" for *Usage (WoS)* / *Readers (Mendeley)* to *Citations (WoS)*, confirming the observations by Chi and Glänzel (2018) in the context of usage and citations.

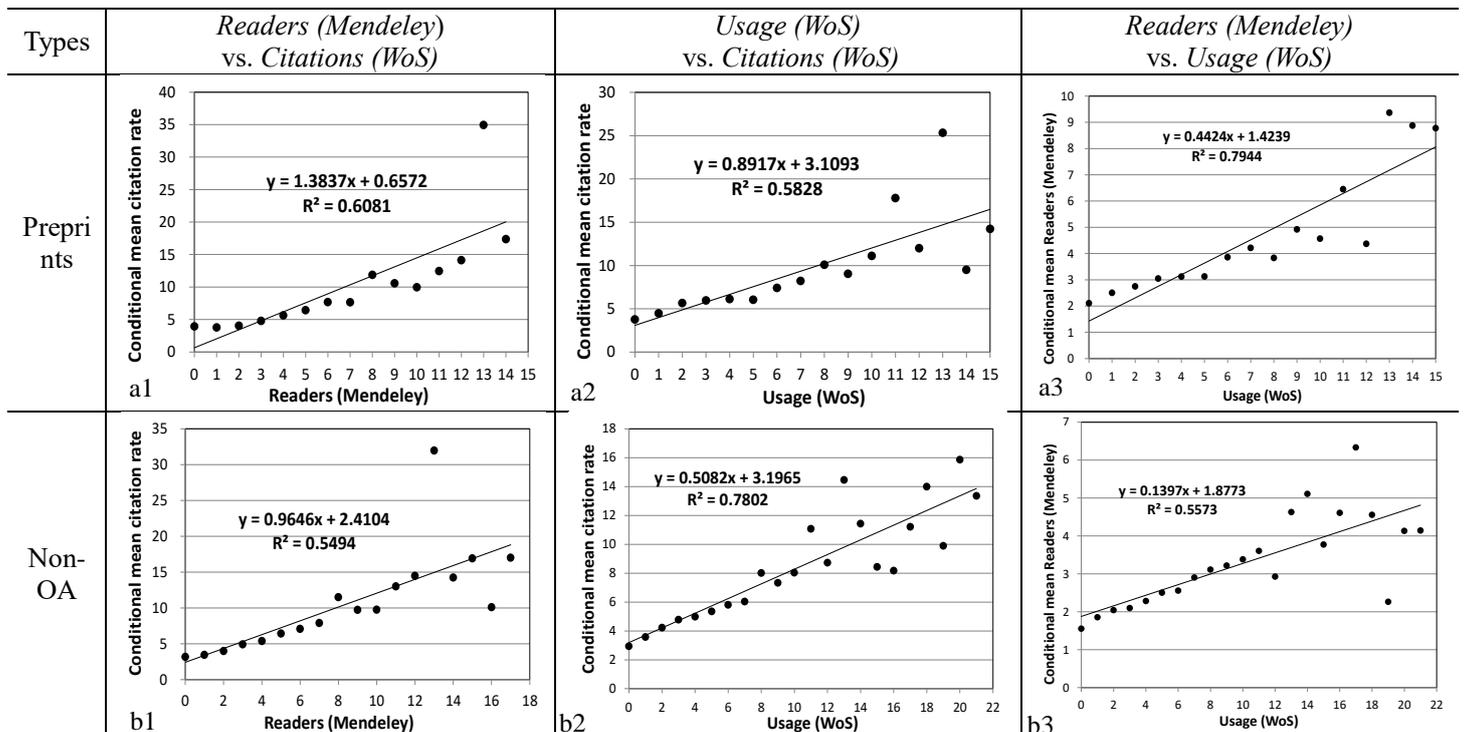

**Fig. 11.** Plots of conditional mean rate of *Citations(WoS)* as a function of *Usage(WoS)* and *Readers(Mendeley)* / conditional mean *Readers(Mendeley)* as a function of *Usage(WoS)* for preprints (a1-a3) and non-OA papers (b1-b3) (M1)

## 5. Conclusions

By analysing 7,669 arXiv preprints and 15,721 non-OA articles published in 220 Mathematics journals indexed in WoS in the publication year 2013, we study the key driving factors of preprints in speeding up scholarly communication from two perspectives, publication delay and impact. Preprint publishing speeds up dissemination of knowledge through the immediacy of sharing research findings with the respective scientific community. The mean publication delay of preprints, i.e., the average time elapsed from arXiv submission date to the journal online date, of the three sub-disciplines in Mathematics amounts to 16.13 months, which is longer than that in Physics or Biology (cf. Larivière et al, 2014; Abdill & Blekhman, 2019). Nearly 94% of all published arXiv papers are posted on arXiv before their first publication in journals. In particular, more than two thirds of the preprints published in the top 16 journals with the largest amount of arXiv papers (Table 2) are submitted to arXiv before being submitted to journals (cf. Fig. 4a). Additionally, the very short time lag between arXiv submission and journal received date indicates that a typical function of preprints is to bridge the publication delay caused by the peer-review process and make research outputs available to the community as early as possible. Furthermore, about 50% of the authors update their preprints with new versions during the ongoing peer-review process after submitting the manuscripts to journals. The latest update time is close to the journal accepted/online time, provided that there is no embargo time required by publishers. The update policy of arXiv severs on one hand enables authors to immediately share the latest follow-up research findings or peer-reviewed manuscripts accepted by journals, on the other hand, promotes the transparency of research process.

The results of the study give evidences that the fast and free dissemination of research results enabled by arXiv speeds up scholarly and more generally scientific communication and, most notably, reception of results in the field under study. 19.5% of preprints are cited by scientific documents indexed in WoS by their unique arXiv identifiers, and 78.1% of these citations are received *before* online publication in a journal. Authors tend to replace citations to preprints by citations to the "more official" journal versions in the reference lists of their papers in Mathematics as well, which has already been observed in the preprint citation analysis of LIS by Wang et al. (2018). Journal papers with preprint versions enjoy higher citation counts compared with papers in the same journal while without any OA versions available. The higher citation rates in a one-year citation window in two of the three analysed sub-disciplines can be interpreted in the context of the above-mentioned "early-view" effect, but the stable citation advantage of preprints versus non-OA papers over years as shown in Fig. 7a and Fig. 10 indicates that other factors may also contribute, such as the open access advantage suggested by Gargouri et al. (2010) or self-selection bias (or quality bias), that is authors preferentially tend to self-archive their better papers or high-impact authors may tend to deposit their papers in arXiv more often, as suggested in some studies (Kurtz et al., 2005; Henneken et al., 2006; Kurtz & Henneken, 2007; Moed, 2007). While the issue on which one may be the stronger explanation for the citation advantage of journal papers with preprint versions deposited in arXiv in the field of mathematics is worth further exploring and this will be part of future research.

Usage and altmetric metrics offer new potential for measuring the broader impact of preprints within the framework of scholarly communication and beyond. We observe that preprint publication also goes with more frequent capture and usage – with one exception, which applies to non-OA publications, where the *Usage (WoS)*, on average, exceeds that of preprints. We interpret this as an effect of the additional services offered by the WoS database that can be used when no free access is granted to the publications in question. The results of the linear regression analysis suggest that the 'preprint-deposited' status remains an important independent predictor of *Citations, Usage (WoS), Readers (Mendeley)* and *Tweets* even when controlling for the additional factors (see Table 4) that are known to associate with the citation counts and altmetric data. In addition, it is interesting to see that compared with the "more official" journal versions, preprint versions are more likely to be tracked by Altmetric.com, in particular, mentioned on Twitter, with a much higher presence and density of Tweets (see Table 5). Preprints also have shorter *Altmetric attention delay*, i.e., the majority (75.50%) of the preprints covered in Altmetric.com are tracked within five days after being submitted to arXiv and in most cases the Altmetric '*first seen date*' is earlier than the journal online date of a preprint (cf. Fig. 9). Given that the audience on social media platforms such as Twitter are more general public outside academia who may who may lack access to the final journal versions, sharing the openly accessible preprint versions on Twitter may gain more potential readers and is good for facilitating the direct communication between readers and authors. On the other hand, the results give evidence that the social media, in particularly Twitter, plays an increasingly important role in supporting the perception of preprints in scholarly and broader communication.

The multiple indicators related to the citation, usage, capture and social media attention impact of preprints and non-OA papers in Mathematics are compared and Spearman correlations between each other are reported. *Readers (Mendeley)* and *Usage (WoS)* both have moderate but stronger correlations with *Citations* than those between *Tweets* and *Citations* in both data sets (i.e. preprints and non-OA papers) across three sub-disciplines. Given the highest *IDR* values of *Tweets (a-ID)* of preprints (cf. Fig. 10) , the low correlations between number of *Tweets* and *citations* (cf. Table 6) indicate that Tweets-based indicators capture other facets of impact of preprints that can be hard to be detected in citations. In addition, preprints are found slightly higher correlation of *Readers (Mendeley)* vs. *Citations (WoS / SC)* than that of *Usage (WoS)* vs. *Citations (WoS / SC)*. The results of the correlations presented in Table 6 give evidence that Mendeley readers may be the most promising alternative indicators to citation-based research evaluation of preprints impact. Finally, the different observed slopes regarding the regression lines between *Readers (Mendeley)* and *Usage (WoS)* vs. *Citations (WoS)* and between the two "document types" express the specific "translation factors" for usage counts in WoS, Mendeley readers and citations in WoS (cf. Chi & Glänzel, 2018).

However, the correlations and regressions between the altmetric indicators and citation indicators do not imply any causal relationships. We expect the added value of the new metrics in reflecting the wider impact of preprints which beyond the framework of traditional bibliometric indicators. We also acknowledge that there are some limitations in the study, the first is that we do not control the effect of the self-selection bias (or quality bias) of arXiv papers on citations, and therefore our findings cannot infer the causal effect of preprint on citations, usage, capture and social attention indicators. Besides the effects of the nature of preprints, there are other additional factors, such as the journal impact factor, number of co-authors, references and pages (Gargouri et al., 2010), authors' influence (Feldman, Lo, & Ammar, 2018), academic age, country, institution and gender (Fraser et al., 2020), would also influence the impact indicator differentials between arXiv and non-arXiv papers. A recent study (Fraser et al., 2020) conducted a negative binomial regression analysis to investigate the influence of additional factors on citation and altmetric differentials between bioRxiv-deposited and non-deposited papers, and the results showed that when controlling for a set of explanatory variables related to publication venue and authorship, the bioRxiv deposit status remains an important independent predictor of citations and altmetric indicators. However, as stressed by the authors (Fraser et al., 2020), these results still cannot establish the causal relationship, since the factors that can influence a paper's citation or altmetric counts are very complicated, and some of them are immeasurable variables such as a paper's underlying quality or the author' selection bias on preprint posting. Future research could do more on decomposing these effects and measuring direct effects of the individual factors that may influence an article's citation or alternative metrics, which would help provide deeper insight into the driving factors of preprints as accelerator of scholarly communication.

This study provides a new perspective to analyse the role of preprints in scholarly and broader scientific communication, while many questions still remain to be answered, for example, how preprints are cited in preprints or other non-journal publications including thesis, books, conference proceedings, reports and web pages, and how and what the difference between knowledge transfer patterns in preprints and in traditional journal publications might be. These will be part of our future research.

# Appendix

**Table A.** Results of the influence of the full variables for the citations, usage, capture and social media attention indicators

| Independent Variables | Citations (WoS) | | | Citations (SC) | | | Usage (WoS) | | |
|---|---|---|---|---|---|---|---|---|---|
| | $\beta_{full\ model}$ (95%CI) | Std. Error | ρ | $\beta_{full\ model}$ (95%CI) | Std. Error | ρ | $\beta_{full\ model}$ (95%CI) | Std. Error | ρ |
| Constant | -0.647 (-0.733 - -0.560) | 0.044 | 0.000 | -0.493 (-0.582 - -0.404) | 0.045 | 0.000 | 0.379 (0.303 - 0.456) | 0.039 | 0.000 |
| **Preprint-deposited** | 0.138 (0.113 - 0.163) | 0.013 | 0.000 | 0.149 (0.123 - 0.174) | 0.013 | 0.000 | -0.218 (-0.241 - -0.196) | 0.011 | 0.000 |
| Journal Impact factor | 0.486 (0.461 - 0.510) | 0.012 | 0.000 | 0.501 (0.475 - 0.526) | 0.013 | 0.000 | 0.375 (0.353 - 0.396) | 0.011 | 0.000 |
| Review article | 0.081 (-0.135 - 0.297) | 0.110 | 0.460 | 0.037 (-0.186 - 0.259) | 0.113 | 0.747 | 0.47 (0.278 - 0.662) | 0.098 | 0.000 |
| First author academic age* | 0.008 (-0.005 - 0.021) | 0.007 | 0.232 | 0.005 (-0.008 - 0.019) | 0.007 | 0.418 | -0.045 (-0.056 - -0.034) | 0.006 | 0.000 |
| Last author academic age* | 0.018 (0.005 - 0.031) | 0.007 | 0.007 | 0.019 (0.005 - 0.032) | 0.007 | 0.006 | 0.002 (-0.010 - 0.013) | 0.006 | 0.790 |
| First author is from USA | -0.008 (-0.049 - 0.032) | 0.021 | 0.685 | 0.009 (-0.033 - 0.050) | 0.021 | 0.684 | -0.13 (-0.166 - -0.095) | 0.018 | 0.000 |
| Last author is from USA | -0.004 (-0.044 - 0.035) | 0.020 | 0.832 | -0.008 (-0.049 - 0.032) | 0.021 | 0.684 | -0.012 (-0.047 - 0.023) | 0.018 | 0.496 |
| Two Authors | 0.143 (0.116 - 0.170) | 0.014 | 0.000 | 0.173 (0.145 - 0.201) | 0.014 | 0.000 | 0.227 (0.203 - 0.251) | 0.012 | 0.000 |
| Three Authors | 0.249 (0.217 - 0.281) | 0.016 | 0.000 | 0.285 (0.252 - 0.317) | 0.017 | 0.000 | 0.376 (0.348 - 0.404) | 0.014 | 0.000 |
| Four Authors | 0.313 (0.264 - 0.361) | 0.025 | 0.000 | 0.351 (0.301 - 0.401) | 0.026 | 0.000 | 0.465 (0.422 - 0.508) | 0.022 | 0.000 |
| Five Authors or more | 0.402 (0.329 - 0.474) | 0.037 | 0.000 | 0.463 (0.388 - 0.538) | 0.038 | 0.000 | 0.631 (0.566 - 0.695) | 0.033 | 0.000 |
| Number of References* | 0.359 (0.338 - 0.380) | 0.011 | 0.000 | 0.368 (0.346 - 0.389) | 0.011 | 0.000 | 0.409 (0.390 - 0.427) | 0.009 | 0.000 |
| Number of pages* | 0.083 (0.061 - 0.106) | 0.012 | 0.000 | 0.059 (0.036 - 0.082) | 0.012 | 0.000 | -0.251 (-0.271 - -0.231) | 0.010 | 0.000 |
| MP | -0.016 (-0.048 - 0.017) | 0.016 | 0.345 | -0.101 (-0.134 - -0.067) | 0.017 | 0.000 | -0.273 (-0.302 - -0.244) | 0.015 | 0.000 |
| MA | 0.03 (0.003 - 0.057) | 0.014 | 0.030 | 0.011 (-0.017 - 0.039) | 0.014 | 0.441 | 0.189 (0.165 - 0.213) | 0.012 | 0.000 |
| MIA | -0.141 (-0.186 - -0.096) | 0.023 | 0.000 | -0.222 (-0.268 - -0.176) | 0.024 | 0.000 | 0.623 (0.583 - 0.663) | 0.020 | 0.000 |

| Independent Variables | Readers (Mendeley) | | | Tweets (a-DOI) | | | Tweets (a-ID) | | |
|---|---|---|---|---|---|---|---|---|---|
| | $\beta_{full\ model}$ (95%CI) | Std. Error | ρ | $\beta_{full\ model}$ (95%CI) | Std. Error | ρ | $\beta_{full\ model}$ (95%CI) | Std. Error | ρ |
| Constant | 0.155 (0.073 - 0.237) | 0.042 | 0.000 | 0.012 (-0.005 - 0.029) | 0.009 | 0.159 | 0.712 (0.585 - 0.839) | 0.065 | 0.000 |
| **Preprint-deposited** | 0.249 (0.225 - 0.272) | 0.012 | 0.000 | 0.033 (0.028 - 0.038) | 0.003 | 0.000 | 0.133 (0.096 - 0.17) | 0.019 | 0.000 |
| Journal Impact factor | 0.334 (0.311 - 0.357) | 0.012 | 0.000 | -0.002 (-0.007 - 0.003) | 0.002 | 0.411 | -0.033 (-0.069 - 0.003) | 0.018 | 0.071 |
| Review article | 0.33 (0.126 - 0.534) | 0.104 | 0.002 | 0.108 (0.066 - 0.150) | 0.021 | 0.000 | 0.622 (0.303 - 0.940) | 0.162 | 0.000 |
| First author academic age* | 0.007 (-0.005 - 0.019) | 0.006 | 0.256 | -0.003 (-0.006 - -0.001) | 0.001 | 0.015 | 0.008 (-0.011 - 0.026) | 0.010 | 0.429 |
| Last author academic age* | 0.044 (0.032 - 0.056) | 0.006 | 0.000 | 0.002 (0.000 - 0.005) | 0.001 | 0.055 | -0.007 (-0.026 - 0.012) | 0.010 | 0.442 |
| First author is from USA | 0.097 (0.058 - 0.135) | 0.019 | 0.000 | 0.009 (0.001 - 0.017) | 0.004 | 0.028 | -0.051 (-0.111 - 0.008) | 0.030 | 0.089 |
| Last author is from USA | 0.082 (0.045 - 0.120) | 0.019 | 0.000 | 0.004 (-0.004 - 0.012) | 0.004 | 0.345 | 0.049 (-0.010 - 0.107) | 0.030 | 0.101 |
| Two Authors | 0.123 (0.097 - 0.149) | 0.013 | 0.000 | -0.004 (-0.010 - 0.001) | 0.003 | 0.121 | -0.038 (-0.078 - 0.002) | 0.020 | 0.061 |
| Three Author | 0.226 (0.196 - 0.256) | 0.015 | 0.000 | -0.004 (-0.010 - 0.002) | 0.003 | 0.225 | -0.047 (-0.094 - 0.000) | 0.024 | 0.048 |
| Four Author | 0.383 (0.336 - 0.429) | 0.024 | 0.000 | 0.009 (0.000 - 0.019) | 0.005 | 0.055 | 0.006 (-0.066 - 0.078) | 0.037 | 0.872 |
| Five Authors or more | 0.542 (0.473 - 0.611) | 0.035 | 0.000 | 0.041 (0.027 - 0.055) | 0.007 | 0.000 | -0.009 (-0.116 - 0.098) | 0.054 | 0.870 |
| Number of References* | 0.228 (0.208 - 0.247) | 0.010 | 0.000 | -0.018 (-0.023 - -0.014) | 0.002 | 0.000 | 0.07 (0.040 - 0.101) | 0.015 | 0.000 |
| Number of pages* | -0.108 (-0.129 - -0.086) | 0.011 | 0.000 | 0.024 (0.019 - 0.028) | 0.002 | 0.000 | -0.078 (-0.112 - -0.045) | 0.017 | 0.000 |
| MP | -0.284 (-0.314 - -0.253) | 0.016 | 0.000 | -0.021 (-0.028 - -0.015) | 0.003 | 0.000 | -0.025 (-0.073 - 0.022) | 0.024 | 0.297 |
| MA | -0.073 (-0.098 - -0.047) | 0.013 | 0.000 | -0.008 (-0.014 - -0.003) | 0.003 | 0.002 | -0.007 (-0.047 - 0.033) | 0.02 | 0.726 |
| MIA | 0.109 (0.066 - 0.151) | 0.022 | 0.000 | 0.045 (0.036 - 0.054) | 0.005 | 0.000 | 0.154 (0.088 - 0.220) | 0.034 | 0.000 |

Note: *Log Transformed